\documentclass{Interspeech2024}

\interspeechcameraready

\usepackage{tcolorbox}
\usepackage{multirow}
\usepackage{makecell}
\usepackage{multicol, subcaption, balance} 
\usepackage{mathtools, nccmath}
\usepackage{appendix}

\title{Spoof Diarization: ``What Spoofed When'' in Partially Spoofed Audio}

\name[affiliation={1,2}]{Lin}{Zhang}
\name[affiliation={1}]{Xin}{Wang}
\name[affiliation={1,3}]{Erica}{Cooper}
\name[affiliation={4}]{Mireia}{Diez}
\name[affiliation={4}]{Federico}{Landini}
\name[affiliation={5}]{Nicholas}{Evans}
\name[affiliation={1,2}]{Junichi}{Yamagishi}

\address{
  $^1$National Institute of Informatics, Tokyo, Japan  $^2$SOKENDAI, Kanagawa, Japan \\
  $^3$National Institute of Information and Communications Technology, Kyoto, Japan \\
  $^4$Brno University of Technology, Faculty of Information Technology, Speech@FIT, Czechia \\
  $^5$Digital Security Department, EURECOM, France }
\email{\{partialspoof, lzhang.as\}@gmail.com \{wangxin, jyamagis\}@nii.ac.jp}


\begin{document}

\maketitle

\begin{abstract}

This paper defines \textbf{Spoof Diarization} as a novel task in the Partial Spoof (PS) scenario. 
It aims to determine \textit{what spoofed when,} which includes not only locating spoof regions but also clustering them according to different spoofing methods. As a pioneering study in spoof diarization, we focus on defining the task, establishing evaluation metrics, and proposing a benchmark model, namely the Countermeasure-Condition Clustering (3C) model. 
Utilizing this model, we first explore how to effectively train countermeasures to support spoof diarization using three labeling schemes.
We then utilize spoof localization predictions to enhance the diarization performance. 
This first study reveals the high complexity of the task, even in restricted scenarios where only a single speaker per audio file and an oracle number of spoofing methods are considered.
Our code is available at \url{https://github.com/nii-yamagishilab/PartialSpoof}.

\end{abstract}

\section{Introduction}

The Partial Spoof (PS) scenario has recently drawn increasing attention \cite{zhang2022partialspoof, yi2022add}. In the conventional fully spoofed scenario, the entire audio signals -- typically an \textit{utterance} within the speech spoofing community -- are generated through Text-to-speech (TTS) and/or Voice Conversion (VC) algorithms \cite{Wang2020data, asvspoof2021}.
In the PS scenario, partially spoofed audio contains segments
generated by TTS/VC, with the remaining regions originating from real human speech \cite{zhang2022partialspoof}, 
where all these segments can have varying durations. 
Compared with the conventional fully spoofed scenario, the PS scenario is more realistic and threatening. Attackers may not need to construct an entire spoofed audio to achieve their goals. Instead, it is more efficient to manipulate only a few arbitrary, short parts of an audio file to drastically distort its original meaning with phonology knowledge and advanced generative models \cite{Zhang2021PartialSpoof}.

A number of studies have addressed the PS scenario. Some investigate spoof detection~\cite{yi2022add, Zhang2021PartialSpoof, martin2022add, wu2022add} to detect whether an audio is spoofed, while others explored spoof localization~\cite{Yi2021halftruth, zhang21partialspoof_mtl, zhang2022localizing} to identify specific suspect segments within an otherwise bona fide audio file. 
However, these two tasks only focus on the binary (yes/no) question to answer whether an audio or segment is spoofed. Thus, solutions for these tasks may be insufficient for realistic forensic scenarios when it is crucial to know not only whether the audio is spoofed but also to obtain detailed information about spoofed segments, like specific spoofing methods or other cues from the spoofing algorithms. This information could potentially aid in tracing the origin or creator of the spoof. We call this task ``Spoof Diarization.'' When the partially spoofed audio contains segments that are created using multiple generative models or through a recursive generation process, spoof diarization can help with traceability, which is crucial in a court of law. Similar investigations have been conducted in the tampering field to trace devices used for recording \cite{cuccovillo2013tamperingdevice, leonzio2023audio}. In contrast, binary-classification-based spoof detection and localization lack direct traceability for such purposes. 
Therefore, it is essential to explore the question of \textit{what spoofed when.}

Speaker diarization \cite{ anguera2012review-sd, park2022review-sd} is a popular task in the speech processing field, aiming to determine ``who spoke when'' for a given recording. Each recording usually involves an unknown number of speakers whose speech duration varies, as seen in interviews, meetings, broadcasts, etc. In speaker diarization, the input audio file is divided into speaker-homogeneous regions (turns), which are clustered into different groups according to speaker characteristics in those regions. Note that speakers' turns might overlap, with systems expected to correctly label such situations as well.

Aligned with speaker diarization, we define the task ``Spoof Diarization'' which involves not only locating spoof regions but also clustering them according to spoofing methods. 
Unlike speaker diarization where clustering is performed according to speaker characteristics, clustering in spoof diarization depends on the spoofing methods. Meanwhile, spoof diarization considers two primary groups of clusters: bona fide and spoof.

In this paper, we focus on defining the task and its evaluation metrics. Further, we present a benchmark model and analyze different labeling schemes corresponding to the specificities of the task.

\section{Spoof Diarization}\label{sec:spfdia}

\begin{figure}[!t]
	\centerline{\includegraphics[width=\linewidth]{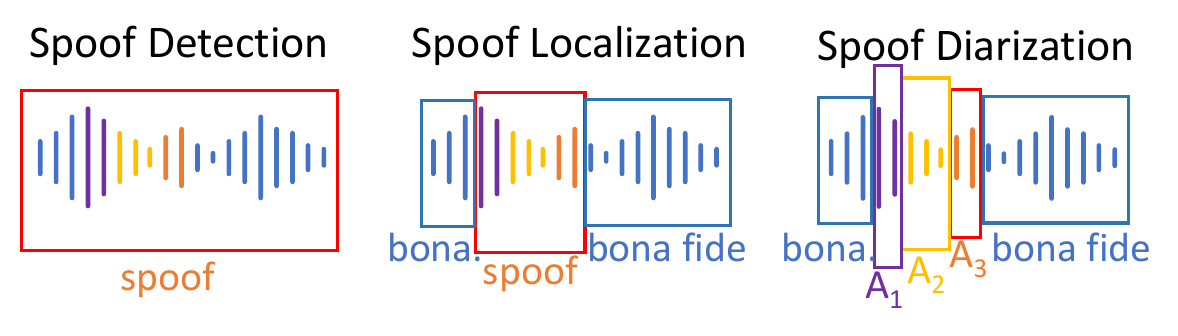}}
  \vspace{-5mm}
\caption{Spoof detection, localization, and diarization.} \label{fig:spfdec_loc_dia}
 \vspace{-5mm}
\end{figure}

\subsection{Definition}\label{sec:define}

\begin{figure}[!t]
 \centering
 \includegraphics[width=\linewidth]{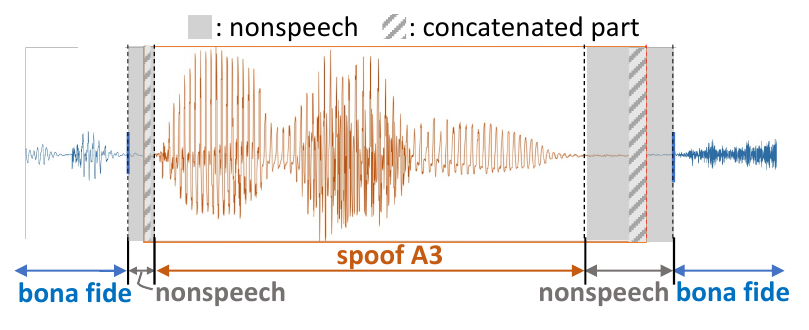}
  \vspace*{-4mm}
\caption{Example of annotated class-homogeneous segments within an audio in the PS scenario.} \label{fig:ConP}
\vspace{-2mm}
\end{figure}

To ``diarize'' means to make a note or keep track of an event in a diary \cite{anguera2012review-sd}. ``Spoof'' indicates falsely claiming a speaker identity \cite{Hadid15_spoof}. Here we define ``Spoof Diarization'' to annotate the spoofing events. Spoof diarization aims to address \textit{``what spoofed when''} in an audio that contains an unknown number of spoofed segments with variable durations. It is related to but different from two well-known tasks in the PS scenario, spoof detection \cite{yi2022add, Zhang2021PartialSpoof} and spoof localization \cite{zhang2022partialspoof, Yi2021halftruth, zhang2022localizing}, as shown in Fig.~\ref{fig:spfdec_loc_dia}. 
Spoof detection aims to determine whether an audio contains any spoofed segment. Spoof localization aims to locate spoof and bona fide regions within an audio. Spoof diarization can be viewed as an extension of spoof localization, where segments generated by different spoofing methods are distinguished and assigned different labels. It can be formulated as:

\begin{tcolorbox}

\begin{itemize}
\item \textbf{Spoof Diarization:} learn a function $f_\text{dia}$ that takes an audio input $\boldsymbol{x}_{1:T}$, and produces a sequence of multi-class labels $\boldsymbol{c}_{1:M}$:
\end{itemize}
$$
\begin{aligned}
\begin{gathered}
f_\text{dia}: \phantom{x}\boldsymbol{x}_{1:T}\mapsto \boldsymbol{c}_{1:M},\\
c_m \in \{bona\phantom{x}fide, A_1,\cdots, A_N, [ConP] \}.
\end{gathered}
\end{aligned}
$$
\end{tcolorbox}
\noindent Here, $\boldsymbol{x}_{1:T}$ denotes a waveform with $T$ samples, and $\boldsymbol{c}_{1:M}$ denotes the segment-level predictions for $M$ segments. Segments may vary or be uniform in duration, depending on the model, with the minimum duration being a frame (as in this paper). 
Fig.~\ref{fig:ConP} shows an example annotation. $A_*$ denotes different types of spoofing methods. Note that,
during prediction, $N$ is expected to be unknown in real-world scenarios, with a majority of $A_*$ unseen in training data, posing an ``open-set'' challenge.
$ConP$ represents concatenated parts where segments with different classes are seamlessly joined. It can be implemented by signal-processing techniques \cite{zhang2022partialspoof, Yi2021halftruth}, neural network-based approaches \cite{cai2023avdf1m}, etc., which could also introduce artifacts and could be treated as a special type of spoof. Whether to include it depends on the data design and model implementation.

The ideal spoof diarization system should be able to classify spoofing methods seen in the train set and group unseen methods as in object detection and discovery \cite{Zheng2022OSODD}. As an initial study on this topic, we cluster (without identifying) all segments as done in speaker diarization, and all $A_*$, seen or unseen during training, will be considered independently on evaluation. Addressing the identification of the exact spoofing methods, especially in an open-set scenario, is designated for future work.

\subsection{Spoof diarization and speaker diarization}\label{sec:spfdia_spkdia}
\begin{figure}[!t]
     \captionsetup[subfigure]{justification=centering}
     \centering
    \begin{subfigure}[!t]{0.13\textwidth}
        \centering
         \includegraphics[width=\textwidth]{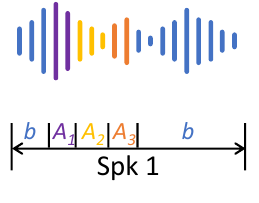}
       \vspace{-2mm}
        \caption{ \\ Spoof Diarization}
        \label{fig:spfdia}
    \end{subfigure}
    \hfill
    \begin{subfigure}[!t]{0.15\textwidth}
        \centering
        \includegraphics[width=\textwidth]{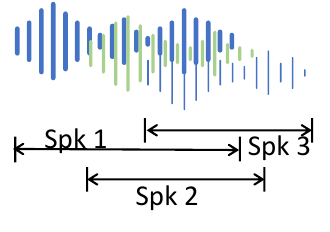}
        \vspace{-3mm}
        \caption{\\Speaker Diarization}\label{fig:spkdia}
    \end{subfigure}
    \hfill
    \begin{subfigure}[!t]{0.18\textwidth}
        \centering
        \includegraphics[width=\textwidth]{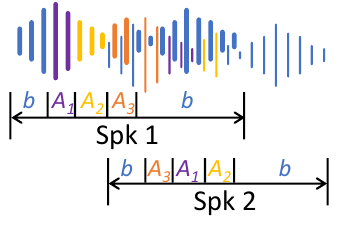}
        \vspace{-7mm}
        \caption{\\ Spoof-Speaker Diarization}\label{fig:spfspkdia}
    \end{subfigure}
 \vspace{-3mm}
 	\caption{Comparison of different diarization tasks. ``b'' for bona fide,``$A_*$'' for spoofing methods, and ``Spk*'' for speakers. Nonspeech is omitted for clarity.} \label{fig:spfdia_spkdia}
   \vspace{-3mm}
\end{figure}

In the speech processing field, one of the well-known diarization problems is speaker diarization, which aims to determine ``who spoke when.''  
A comparison of Fig.~\ref{fig:spfdia} (spoof diarization) and \ref{fig:spkdia} (speaker diarization) shows similarities and differences between the two tasks:

\begin{itemize}[leftmargin=*, itemsep=5mm, nosep]
\item\textbf{Similarities:}
\begin{enumerate}[itemsep=0mm, nosep]
    \item Audio samples in both spoof diarization and speaker diarization are generated by an unknown number of classes (spoofing methods or speakers),
    \item Class-homogeneous regions in both tasks can have variable durations.
\end{enumerate}    

\item \textbf{Differences:}
\begin{enumerate}[itemsep=0mm, nosep]
    \item \textit{Duration of turns}: 
    The relevance of detecting short (word level) turns in speaker diarization systems depends on the application and is not relevant or even not evaluated for some of them. In contrast, the detection of such common short-turn spoofed speech (a single word or even a single phoneme) is crucial, as it can completely change the meaning of the audio. e.g., ``lock account'' to ``unlock account.''
    
    \item \textit{Two primary groups of clusters}: In speaker diarization, speakers may vary for each audio. However, in spoof diarization, two primary groups should be considered: bona fide and spoof.

\end{enumerate}

\end{itemize}

Spoof-speaker diarization, as shown in Fig.~\ref{fig:spfspkdia}, extends spoof diarization to the case of an audio with multiple speakers. This study focuses on spoof diarization, and multi-speaker partially spoofed audio is left for future work.

\subsection{Metric - Spoof Jaccard error rate}\label{sec:JER}
In speaker diarization, there are two main metrics: diarization error rate (DER) \cite{nist2006der}, and Jaccard error rate (JER) \cite{Ryant2019-dihard2}.
While DER accounts for all errors with respect to the total duration of speech in recordings, JER gives equal weight to all speakers, independently of their relative activity. Given the nature of spoof diarization, where a very short segment of spoofed speech can have a large impact, the JER metric matches the task better.

Spoof diarization has two important goals: (1) differentiating spoofed from bona fide segments, and (2) discriminating different spoofing methods. 
To measure the performances regarding these two goals, we adapted the $\mathtt{JER}$\footnote{\url{https://github.com/nryant/dscore}} with two forms: $\mathtt{JI}_\text{bona}$\footnote{JI is used instead of JER because it only considers a single class - bona fide, which aligns better with the Jaccard Index.} and $\mathtt{JER}_\text{spoof}$, respectively. 
\begin{figure*}[!thb]
    \centering
    \begin{subfigure}[b]{0.31\textwidth}
        \centering
         \includegraphics[width=.95\textwidth]{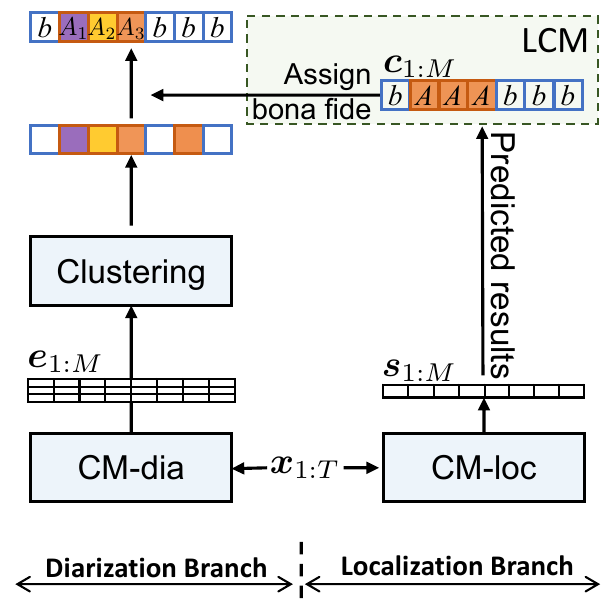}
       \vspace{-2mm}
        \caption{3C model with two branches.}
        \label{fig:cm_cluster}
    \end{subfigure}
    \hfill
    \begin{subfigure}[b]{0.24\textwidth}
        \centering
        \includegraphics[width=.95\textwidth]{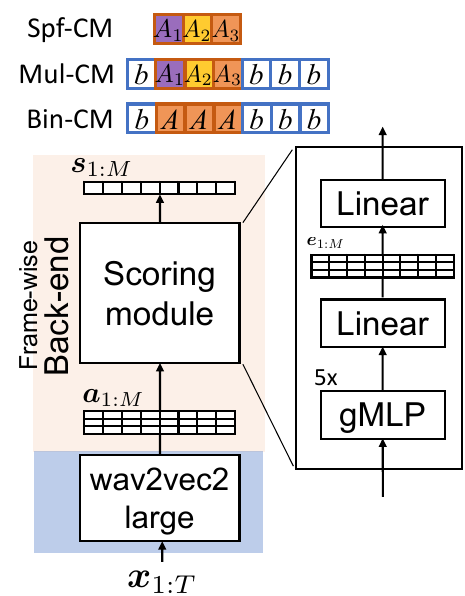}
        \vspace{-2mm}
        \caption{Structure of CMs in (a).}\label{fig:CM}
    \end{subfigure}
    \hfill\hfill
    \begin{subfigure}[b]{0.42\textwidth}
        \centering
        \includegraphics[width=.95\textwidth]{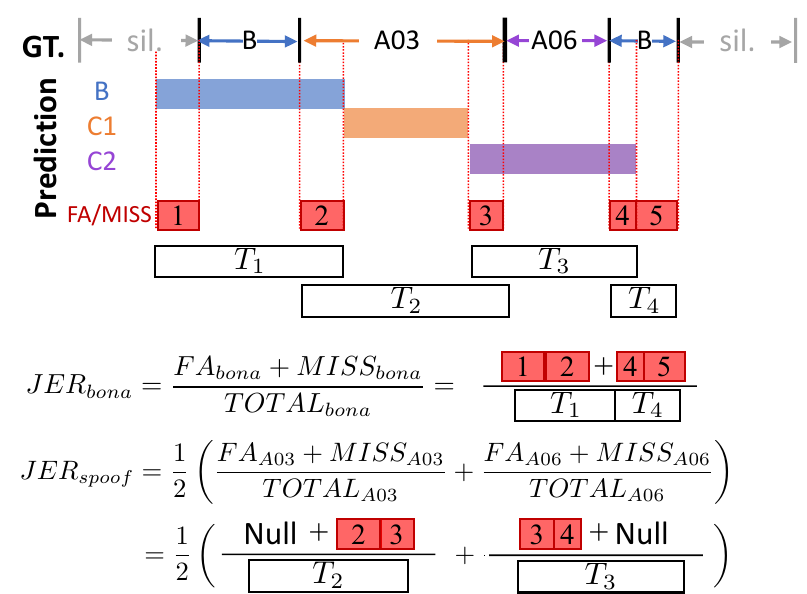}
        \vspace{-2mm}
        \caption{$\mathtt{JI}_\text{bona}$ and $\mathtt{JER}_\text{spoof}$ for spoof diarization.}\label{fig:JER_spoof_dia}
    \end{subfigure}
    \vspace{-2mm}
    \caption{Proposed benchmark model and metrics for spoof diarization.}
   \label{fig:model_metric}
   \vspace{-4mm}
\end{figure*}

They are calculated after an optimal mapping between the reference class and the predicted cluster. This mapping can be determined by the Hungarian algorithm \cite{kuhn1955hungarian} following the common approach in speaker diarization \cite{pyannote.metrics, nist2009der}.
An example after best mapping is shown in Fig.~\ref{fig:JER_spoof_dia}. 
$\mathtt{JI}_\text{bona}$ and $\mathtt{JER}_\text{spoof}$ for the $j$-th audio are given by:
\vspace{-0.06 in}
\begin{equation}
    \begin{aligned}
    \mathtt{JI}_{\text{bona}, j} & = \frac{\mathtt{FA}_{\text{bona}, j}+\mathtt{MD}_{\text{bona}, j}}{\mathtt{TOTAL}_{\text{bona}, j}},
     \end{aligned}
\end{equation}   
    \begin{equation}
    \begin{aligned}
    \mathtt{JER}_{\text{spoof}, j} & = \frac{1}{|\mathcal{A}_j|}\sum_{A_i\in \mathcal{A}_j}\mathtt{JER}_{A_i, j} = \frac{1}{|\mathcal{A}_j|} \sum_{A_i \in \mathcal{A}_j} \frac{\mathtt{FA}_{A_i, j}+\mathtt{MD}_{A_i, j}}{\mathtt{TOTAL}_{A_i, j}}.
    \end{aligned}
\end{equation}
Where $\mathtt{FA}_{j}$ and $\mathtt{MD}_{j}$ present false alarms and missed detections in the $j$-th trial, respectively. $\mathtt{TOTAL}_j$ refers to the union duration between reference and prediction.
Subscripts $_\text{bona}$ and $_{A_i}$ indicate bona fide and a specific spoofing method $A_i$, respectively.
$\mathcal{A}_j$ is the set of different spoofing methods within the $j$-th audio, and $|\cdot|$ denotes the size of the set.

For the global evaluation of a set of audio files $\mathcal{D}$, we use macro averaging: 
\vspace{-0.12 in}
\begin{equation}
    \begin{aligned}
    \mathtt{JI}_{\text{global}\_\text{bona}} = \frac{1}{|\mathcal{D}|}\sum_{j \in \mathcal{D}} \mathtt{JI}_{\text{bona}, j}
    \end{aligned}
\end{equation}   
\vspace{-2.5mm}
\begin{equation}
    \begin{aligned}
    \mathtt{JER}_{\text{global}\_\text{spoof}} = \frac{1}{\sum_{j \in \mathcal{D}}|\mathcal{A}_j|} \sum_{j \in \mathcal{D}}\sum_{A_i \in \mathcal{A}_j}\mathtt{JER}_{A_{i}, j}
    \end{aligned}
\end{equation}  
To simplify the notation, we drop $_\text{global}$ from the subscript in the rest of this paper, e.g., $\mathtt{JI}_{\text{global}\_\text{bona}}$ becomes $\mathtt{JI}_\text{bona}$.

\section{Proposed 3C Model for Spoof Diarization}\label{sec:model_3C}

\subsection{3C model: CM-condition clustering}

The proposed 3C model is depicted in Fig.~\ref{fig:cm_cluster}, with the structure of its CMs shown in (b). The 3C model comprises two branches: (1) The diarization branch parallels the conventional modular-based speaker diarization pipeline. It consists of an embedding extractor (CM-dia) followed by a clustering. And (2) the localization branch provides additional bona fide information derived from CM-loc to the diarization branch. CMs in both branches use frame-wise uniform segmentation within audio to facilitate accurate diarization.

To formulate the processing, given an input $\boldsymbol{x}_{1:T}$, we first extract embeddings $\boldsymbol{e}_{1:M}$ from CM-dia. Following this, we calculate a pairwise affinity matrix which will be used for further clustering. 
From CM-loc, we derive frame-level predicted scores $\boldsymbol{s}_{1:M}$ to determine bona fide frames. It can be implemented by either binary classification or multi-classification strategies. For the binary classification, frame-wise equal error rate \cite{zhang23rangeEER} is estimated on the development set, and its corresponding threshold $\tau$ is used to classify the $i$-th frame as bona fide if $s_i > \tau$, otherwise as spoofed.  
In the case of a multi-classification, where training labels include bona fide and various spoofing methods, the class receiving the highest predicted probability is determined for each frame.
The bona fide predictions obtained with binary or multi-class CM-loc are then used to potentially enhance the accuracy of bona fide prediction.
We use a Label-based CM-constraint (LCM) approach: if CM-loc identifies a frame as bona fide, the final output for this frame is bona fide, (conditioning the CM-dia output).
We also explored the potential of using predicted scores to refine the pairwise affinity matrix, but no notable improvement compared with LCM was observed.

\subsection{Labeling scheme}\label{sec:label}

As introduced in Section \ref{sec:spfdia_spkdia}, in spoof diarization we aim to distinguish bona fide and different spoofing methods.
Therefore, the natural choice for training a spoof diarization method, would be to do it for multi-classification considering the labels of all classes. Still, in the PS scenario, it is more relevant to distinguish the two main categories, namely \textit{bona fide} and \textit{spoof}.
Taking this into consideration, and given that the 3C model proposed for the spoof diarization contains two branches, we explored the possibilities of training each of these branches with different labeling schemes:

\begin{enumerate}[itemsep=0mm, leftmargin=4mm, nosep]
    \item \texttt{Bin-CM}: CM trained for binary-classification with labels \\ 
    $y_m \in \{bona\ fide, spoof\}$, 
    \item \texttt{Mul-CM}: CM trained for multi-classification with labels \\
    $y^\ast_m \in \{bona\ fide, A_1, \cdots A_N, ConP\}$,
    \item \texttt{Spf-CM}: CM trained for multi-spoof classification with \\
    $y^\star_m \in \{A_1, \cdots, A_N, ConP\}$.
\end{enumerate}
\noindent Where \texttt{Bin-CM} focuses solely on binary classification between bona fide and spoof, and aggregates all different spoofing methods as a single \textit{spoof} category following previous spoof detection and localization studies. \texttt{Mul-CM} not only discriminates between spoof and bona fide but also identifies different spoofing methods. Finally, \texttt{Spf-CM}, by excluding bona fide data in its training, aims to explore whether such an approach enhances the ability to differentiate among various spoofing methods.

\section{Experiments and Results}\label{sec:conf}
\subsection{Experimental setup}
We used the PartialSpoof\footnote{\url{https://zenodo.org/records/5766198}} database to explore spoof diarization. In this dataset, the proportions of bona fide are 55.3\%, 56.0\%, and 60.7\% for train, development, and evaluation sets, respectively. Furthermore, the presence of different spoofing methods in the PartialSpoof is relatively even and shows a balanced distribution.
We applied an adapted Oracle voice activity detection (VAD). Specifically, we removed the nonspeech parts, apart from concatenated parts\footnote{While these concatenated parts are originally nonspeech regions within the PartialSpoof database, modifications made to them might have introduced artifacts \cite{zhang2022partialspoof}. And such concatenated parts are proven helpful for spoof detection \cite{cai2022ADD2boundary}. Therefore, we kept and treated them as a unique class during training.}, from training. Nonspeech segments are not taken into account when scoring the predictions. Therefore, errors 1 and 5 shown in Fig.~\ref{fig:JER_spoof_dia} will not happen.

Given the necessity for short-duration diarization highlighted in Section \ref{sec:spfdia_spkdia} -- Difference 1, all CMs in this paper were trained at a 20 ms resolution following \cite{zhang2022partialspoof}. They consist of a wav2vec2-large \cite{BaevskiZMA20-w2v2} as the front-end with gMLPs \cite{Liu2021gmlp} as the back-end and achieve the best performance on the spoof localization. For the binary classification CM-loc, 10 ms frame-wise equal error rate \cite{zhang23rangeEER} was calculated to determine $\tau$ for identity whether a frame is bona fide or spoof.
The frame-wise embeddings $\boldsymbol{e}_{1:M}$ and predicted scores $\boldsymbol{s}_{1:M}$ are extracted per 20 ms. The resolution of 20 ms is given by convolutional layers in the wav2vec2-large model.
The embeddings $\boldsymbol{e}_{1:M}$ are extracted from the penultimate layer of the back-end as shown in Fig.~\ref{fig:CM}.
We utilize the widely used Agglomerative Hierarchical Clustering with cosine distance to cluster the extracted embeddings.

Note that in spoof diarization, the number of spoofing method types can be either known or unknown. In this initial study, we cluster until reaching the oracle number of clusters. In the PS scenario, both spoof and bona fide regions can be arbitrarily short and very difficult to discriminate.  This makes accurately estimating the correct number of clusters a complicated task.
Therefore, this study conducts an analysis in a controlled scenario with a known oracle number of clusters. Future research could fruitfully explore efficient and accurate methods for estimating cluster numbers in partially spoofed audio.

\subsection{How to train CMs to support spoof diarization}\label{sec:how}

\subsubsection{How do labeling schemes affect the ability of CMs?}\label{sec:diff_CM}

As introduced in Sec.~\ref{sec:label}, there are three possible labeling schemes to train CMs. In this subsection, we focus on understanding their impact on CMs for spoof diarization. First, we exclude the localization branch from the 3C model and train only the diarization branch based on the CMs using three different labeling schemes: \texttt{Bin-CM}, \texttt{Mul-CM}, and \texttt{Spf-CM}. 
Results are shown in the top part of Table~\ref{tab:res_oracle_cluster}.

Looking at the results on the development set, the model trained with \texttt{Mul-CM} obtains overall best results. This could be expected, as it is the only labeling scheme that covers all the classes. However, it is somewhat surprising to see that it performs similarly to \texttt{Bin-CM} for bona fide localization, even if the latter was specifically trained only for this task.
Regarding the model \texttt{Spf-CM}, one could expect poor performance on for bona fide location, but the model also underperforms in terms of $\mathtt{JER}_\text{spoof}$, revealing that specific treatment of the bona fide class is needed during system training.

The remarkably higher $\mathtt{JI}_\text{bona}$ and $\mathtt{JER}_\text{spoof}$ observed across all three CMs in the evaluation set is understandable, given that the evaluation set of the PartialSpoof database contains seven out of thirteen spoofing methods that are completely unseen during training \cite{Wang2020data}, thereby presenting a more complex challenge.
Focusing on such evaluation set results, we can still see some interesting patterns: the \texttt{Mul-CM} suffers higher degradation in $JI_{bona}$ compared to \texttt{Bin-CM} on the evaluation set, which might indicate overfitting to the development set.

\subsubsection{How do we utilize CMs trained under varying labeling schemes?}\label{sec:res_3C}

Based on the analysis in the previous section, we first chose \texttt{Mul-CM} as CM-dia to produce embeddings for various spoofing methods. Besides, considering that the task of locating bona fide segments is now part of the localization branch, we still considered \texttt{Spf-CM} to generate the embeddings for clustering.
Moreover, considering that both \texttt{Mul-CM} and \texttt{Bin-CM} showed effective performance in locating bona fide regions, we evaluated these two models as potential options for the CM-loc. The bottom of Tab.~\ref{tab:res_oracle_cluster} shows results for our proposed 3C model.

First, we analyze the efficacy of the proposed 3C model. 
The Dia-\texttt{Mul-CM} models with and without localization branch perform similarly on the development set. However, in the evaluation set, we observe a decrease in $\mathtt{JI}_\text{bona}$ but an increase in $\mathtt{JER}_\text{spoof}$. That is, integrating the localization branch allows for better localization capabilities but it negatively impacts the efficiency of diarizing spoofing methods.
When comparing Dia-\texttt{Spf-CM} before and after introducing the localization branch, we notice not only an (expected) remarkable improvement in $\mathtt{JI}_\text{bona}$, but also an improvement in $\mathtt{JER}_\text{spoof}$ in both the development and evaluation sets.
Still, Dia-\texttt{Mul-CM} performs better than Dia-\texttt{Spf-CM}, which reveals that the Dia-\texttt{Mul-CM} model extracts better embeddings for diarizing spoofing methods.
Therefore, model selection would depend on the relevance of these two goals for the specific use case.

Second, comparing the different CM-loc options, results show that using \texttt{Bin-CM} as CM-loc slightly outperforms using \texttt{Mul-CM} as CM-loc, which aligns with the observation from the previous subsection. That is expected as \texttt{Bin-CM} is specialized in distinguishing bona fide from spoof.

As already pointed out, performance on $\mathtt{JER}_\text{spoof}$\footnote{A breakdown of  $\mathtt{JER}_\text{spoof}$ for the individual spoofing methods in the evaluation set can be found on the arXiv version.} is poor on the evaluation set, mainly due to the complexity posed by the open-set scenario on the spoof diarization task. To get a better insight into the impact that unknown spoofing methods had on performance, we analyzed how the best model, namely Dia-\texttt{Mul-CM} + Loc-\texttt{Bin-CM}, performed in terms of $\mathtt{JER}_\text{spoof}$ in known (11.98\%) and unknown (49.02\%) spoofing methods\footnote{Known and unknown are grouped by the presence of their acoustic model and waveform generator in the training dataset, following ASVspoof 2019 LA database \cite{Nautsch2021spoof19}.} separately. Results show a considerable performance gap between them which will be addressed in future work.

\begin{table}[t]
\caption{Results on the PartialSpoof database. Confidence intervals within ``()'' were calculated using the Interspeech official toolkit with the default configuration. ``\texttt{-CM}'' in the first two columns are omitted.}
\vspace{-2mm}
\setlength{\tabcolsep}{3pt} 
\label{tab:res_oracle_cluster}\centering
\scalebox{.91}{
\begin{tabular}{p{6mm}<{\centering}p{6mm}<{\centering}p{16mm}<{\centering}p{16mm}<{\centering}p{16mm}<{\centering}p{16mm}<{\centering}}
\toprule
\multicolumn{2}{c}{Model}     & \multicolumn{2}{c}{Development set} & \multicolumn{2}{c}{Evaluation set} \\
Dia.        & Loc.  & $\mathtt{JI}_\text{bona}$ (\%) & $\mathtt{JER}_\text{spoof}$ (\%)  & $\mathtt{JI}_\text{bona}$ (\%) & $\mathtt{JER}_\text{spoof}$ (\%)     \\
  \midrule 
\texttt{Bin}                  &    /    & \phantom{0}4.37 \scriptsize($\pm$0.06) & 20.45 \scriptsize($\pm$0.34) & 16.85 \scriptsize($\pm$0.14) & 33.13 \scriptsize($\pm$0.22) \\
\texttt{Mul}                  &    /   & \phantom{0}4.49 \scriptsize($\pm$0.07)   & \phantom{0}5.21 \scriptsize($\pm$0.11)  & 19.66 \scriptsize($\pm$0.14) & 28.05 \scriptsize($\pm$0.17) \\
\texttt{Spf}                  &    /    & 26.17 \scriptsize($\pm$0.23)  & 20.85 \scriptsize($\pm$0.25) & 32.30 \scriptsize($\pm$0.16) & 38.51 \scriptsize($\pm$0.17) \\
\hline
\multirow{2}{*}{\texttt{Mul}} & \texttt{Bin}& \phantom{0}4.49   \scriptsize($\pm$0.07) & \phantom{0}5.27 \scriptsize($\pm$0.11) & 15.15 \scriptsize($\pm$0.12) & 34.13 \scriptsize($\pm$0.19) \\
                         & \texttt{Mul} & \phantom{0}4.59 \scriptsize($\pm$0.07)   & \phantom{0}5.31 \scriptsize($\pm$0.10) & 17.08 \scriptsize($\pm$0.12) & 35.34 \scriptsize($\pm$0.19) \\
 \multirow{2}{*}{\texttt{Spf}} & \texttt{Bin} & \phantom{0}4.52 \scriptsize($\pm$0.07)   & \phantom{0}5.71 \scriptsize($\pm$0.12) & 15.18 \scriptsize($\pm$0.11) & 36.03 \scriptsize($\pm$0.19) \\
                        & \texttt{Mul} & \phantom{0}4.62 \scriptsize($\pm$0.08)   & \phantom{0}5.81 \scriptsize($\pm$0.12) & 17.10 \scriptsize($\pm$0.12) & 37.78 \scriptsize($\pm$0.18) \\
\bottomrule
\end{tabular}}
\vspace{-6mm}
\end{table}

\section{Conclusion }
\label{sec:conclusion}

Using generative models to manipulate arbitrary short segments of an audio can drastically change its original meaning as in the Partial Spoof scenario. Locating and tracing back such partially spoofed segments is crucial for speech security, like in a court of law. Thus, spoof diarization, aiming to answer \emph{what spoofed when}, is an essential task that should be explored for the PS scenario.

This study serves as a foundational exploration, presenting the task definition, evaluation metrics, and a benchmark model. Preliminary results in this first study indicate the high complexity of the task, even in controlled scenarios with just a single speaker per audio and a predetermined oracle number of spoofing methods.
We hope that our insights can inspire and encourage future investigations into this task.

\section{Acknowledgements}
This study is partially supported by JST AIP Acceleration Research (JPMJCR24U3), JST CREST Grant (JPMJCR20D3), JST the establishment of university fellowships Grant (JPMJFS2136), SOKENDAI Student Dispatch Program, and MEXT KAKENHI Grants (21H04906). Mireia Diez and Federico Landini from Brno University of Technology were supported by Czech Ministry of Interior projects Nos. VJ01010108 ``ROZKAZ''. Nicholas Evans from EURECOM was supported by the ANR in France (BRUEL project, ANR-22-CE39-0009).

\bibliographystyle{IEEEtran}
\balance
\bibliography{main}
\onecolumn

\begin{appendices}
\section{Appendix}

\subsection{Break-down of $\mathtt{JER}_\text{spoof}$}

To further analyze the diarization model's performance with respect to specific spoofing methods, we break-down $\mathtt{JER}_\text{spoof}$ to a tailored version of it: spoofing-method-specific $\mathtt{JER}_{A_i}$ for each spoofing method $A_i$. This specialized metric requires the computation of performance for each individual spoofing method. It thus enables a detailed assessment of the model's diarization performance for each particular spoofing method $A_i$:
\begin{equation}
    \begin{aligned}
    \mathtt{JER}_{A_i} &= \frac{1}{|\mathcal{D}|}\sum_{j \in \mathcal{D}}\mathtt{JER}_{j, A_i} \\
    \end{aligned}
\end{equation}

Performance analysis also extends to three groups of spoofing methods \cite{Wang2020data}. These groups are named following the ASVspoof 2019 LA database \cite{Nautsch2021spoof19}, based on their presence in the training dataset. The categories include:

\begin{itemize}
    \item Known attacks: Spoofing methods that are identical to those seen in the training set. \\
    A16, A19,
    
    \item Varied attacks: Spoofing methods whose either the acoustic model or waveform generator is identical to those used in the training.\\
    A07, A08, A09, A17,
    
    \item Unknown attacks: Spoofing methods with both the acoustic model or waveform generator not presented during training.\\
    A10, A11, A12, A13, A14, A15, A18.
\end{itemize}

\begin{table}[!hbt]
\centering
\caption{Break-down of $\mathtt{JER}_\text{spoof}$ on the evaluation set of PartialSpoof.}
\begin{tabular}{ccccccccccc}
\toprule
CM-dia & CM-loc & A07   & A08   & A09   & A10   & A11   & A12   & A13   & A14   & A15   \\
\midrule
\texttt{Bin-CM} &  /    & 16.10 & 19.78 & 19.67 & 58.35 & 35.58 & 52.60 & 15.83 & 19.16 & 59.45 \\
\texttt{Mul-CM}&   /    & 13.99 & 11.02 & 16.90 & 57.19 & 32.95 & 53.92 & 16.61 & 14.81 & 59.76 \\
\texttt{Spf-CM} &  /    & 47.81 & 15.50 & 24.37 & 57.95 & 30.14 & 60.88 & 24.93 & 14.81 & 52.22 \\
\midrule
\multirow{2}{*}{\texttt{Mul-CM}} & \texttt{Bin-CM}   & 12.82 & 11.11 & 16.29 & 84.51 & 44.73 & 70.87 & 16.37 & 16.09 & 86.59 \\
       & \texttt{Mul-CM}   & 14.02 & 11.48 & 17.94 & 85.87 & 43.53 & 76.06 & 16.94 & 16.02 & 89.61 \\
\multirow{2}{*}{\texttt{Spf-CM}}  & \texttt{Bin-CM}   & 25.69 & 10.86 & 20.44 & 85.02 & 43.75 & 70.16 & 18.43 & 16.01 & 87.58 \\
       & \texttt{Mul-CM}   & 27.36 & 12.78 & 23.27 & 86.63 & 44.46 & 76.92 & 19.47 & 17.46 & 90.03 \\
\multirow{2}{*}{\texttt{Bin-CM}}    & \texttt{Bin-CM}   & 16.19 & 19.99 & 20.84 & 85.56 & 46.27 & 70.13 & 16.29 & 20.22 & 86.99 \\
       & \texttt{Mul-CM}   & 18.02 & 21.82 & 23.73 & 87.01 & 49.76 & 77.01 & 17.00 & 21.85 & 89.62 \\
\bottomrule
\toprule
CM-dia & CM-loc & A16   & A17   & A18   & A19   & $\mathtt{JI}_{\text{bona}}$ & $\mathtt{JER}_{\text{spoof}}$   & known & varied & unkown \\
\midrule
\texttt{Bin-CM} &   /     & 19.39 & 28.22 & 21.34 & 23.80 & 15.37    & 29.87 & 21.43 & 20.60  & 37.96  \\
\texttt{Mul-CM}&   /     & 11.18 & 21.10 & 18.59 & 11.99 & 17.88    & 26.12 & 11.56 & 15.49  & 36.78  \\
\texttt{Spf-CM} &   /     & 60.85 & 25.97 & 26.33 & 35.61 & 29.70    & 36.75 & 49.17 & 28.36  & 38.35  \\
\midrule
\multirow{2}{*}{\texttt{Mul-CM}} & \texttt{Bin-CM}   & 11.34 & 23.18 & 16.93 & 12.72 & 13.81    & 32.63 & 11.98 & 15.49  & 49.02  \\
       & \texttt{Mul-CM}   & 11.48 & 22.79 & 20.00 & 12.51 & 15.55    & 33.73 & 11.96 & 16.25  & 50.64  \\
\multirow{2}{*}{\texttt{Spf-CM}} & \texttt{Bin-CM}   & 11.98 & 24.61 & 18.58 & 13.96 & 13.84    & 34.52 & 12.89 & 20.13  & 49.50  \\
       & \texttt{Mul-CM}   & 12.02 & 24.70 & 22.49 & 13.79 & 15.57    & 36.35 & 12.84 & 21.84  & 51.95  \\
\multirow{2}{*}{\texttt{Bin-CM}}    & \texttt{Bin-CM}   & 19.95 & 30.38 & 21.99 & 24.87 & 13.81    & 36.90 & 22.22 & 21.45  & 50.55  \\
       & \texttt{Mul-CM}   & 19.72 & 29.94 & 25.98 & 24.26 & 15.49    & 38.88 & 21.82 & 23.08  & 53.43  \\
\bottomrule
\end{tabular}
\end{table}

\end{appendices}

\end{document}